\documentclass{article}

\usepackage{amsbsy,amssymb,latexsym,amsfonts, amsmath,mathrsfs}
\usepackage{graphicx}

\numberwithin{equation}{section}

\newcommand{\beq}{\begin{equation}}
\newcommand{\eeq}{\end{equation}}
\newcommand{\bea}{\begin{eqnarray}}
\newcommand{\eea}{\end{eqnarray}}

\newcommand{\CB}{{\mathcal B}}

\newcommand{\CN}{{\mathcal N}}

\renewcommand{\thefootnote}{\fnsymbol{footnote}}

\begin{document}

\baselineskip 0.7cm

\begin{titlepage}

\setcounter{page}{0}

\renewcommand{\thefootnote}{\fnsymbol{footnote}}

\begin{flushright}
YITP-10-84, IHES/P/10/34\\
\end{flushright}

\vskip 1.35cm

\begin{center}
{\Large \bf 
Seiberg-Witten curve via generalized matrix model 
}

\vskip 1.2cm 

{\normalsize
Kazunobu Maruyoshi$^1$\footnote{maruyosh(at)yukawa.kyoto-u.ac.jp} 
and Futoshi Yagi$^2$\footnote{fyagi(at)ihes.fr}
}

\vskip 0.8cm

{ \it
$^1$Yukawa Institute for Theoretical Physics, Kyoto University, Kyoto 606-8502, Japan \\
$^2$Institut des Hautes $\acute{E}$tudes Scientifiques, 91440, Bures-sur-Yvette, France
}

\end{center}

\vspace{12mm}

\centerline{{\bf Abstract}}

  We study the generalized matrix model which corresponds to the $n$-point 
  toric Virasoro conformal block.
  This describes four-dimensional $\CN=2$ $SU(2)^n$ gauge theory 
  with circular quiver diagram by the AGT relation.
  We first verify that the generalized matrix model is obtained from the
  perturbative calculation of the Liouville correlation function.
  We then derive the Seiberg-Witten curve for $\CN=2$ gauge theory as a
  spectral curve of the generalized matrix model. 

\end{titlepage}
\newpage

\tableofcontents

\section{Introduction}
\label{sec:intro}
  Recently, an interesting conjecture has been proposed in \cite{AGT} that
  there are two equivalent ways for describing
  two M5-branes wrapped on a Riemann surface.
  One of the ways is a four-dimensional $\CN=2$ superconformal $SU(2)$ quiver gauge theory,
  which is constructed depending on the number of genus and punctures of the Riemann surface 
  \cite{Witten, Gaiotto}.
  The other is the Liouville theory defined on that Riemann surface.
  The proposal, which is often called AGT conjecture \cite{AGT}, 
  is that the Nekrasov partition function \cite{Nekrasov} of 
  such quiver gauge theory can be reproduced by
  the $n$-point conformal block of the Virasoro algebra,
  where $n$ is the number of punctures.
  This conjecture has been generalized to the higher rank gauge group 
  \cite{Wyllard, MM, Kanno:2009ga, Kanno:2010kj},
  non-conformal case \cite{Gaiottonon-conf, MMM, Taki:2009zd},
  and to the case in the presence with the surface and loop operators 
  \cite{AGGTV}--\cite{Tai:2010ps}.
  ( See also \cite{Bonelli:2009zp, Alday:2009qq} for the M-theory considerations.)
  A proof has been given in \cite{FL} and \cite{Hadasz:2010xp}
  for $SU(2)$ gauge theory with adjoint and ($N_f=0,1,2$) fundamental hypermultiplets
  by using the recursion relation \cite{Zamolodchikov:1985ie, Marshakov:2009kj, Poghossian, HJS0}.

  It was discussed in \cite{DV} that the AGT conjecture for the case on a sphere 
  can be understood through a matrix model.
  The matrix model expression of the conformal block on a sphere is given 
  by reinterpreting the Dotsenko-Fateev integral representation of it 
  as the (beta-deformed) matrix integral \cite{DF, Marshakov} with a logarithmic potential.
  When the central charge $c = 1 + 6 Q^2$ ($Q = b + 1/b$) equals one, or equivalently, $b=i$, 
  the conformal block is represented by the usual matrix model.
  The matrix model technique, in particular large $N$ limit, is useful to show 
  that the gauge theory result can be reproduced \cite{IMO, EM, Schiappa:2009cc, EM2}.
  (See also \cite{Fujita:2009gf} for $1/N$ correction.)
  It has been discussed that direct integral calculation leads 
  to the instanton ($q$-)expansion of the Nekrasov partition function 
  (and the corresponding expansion of the conformal block)
  \cite{Mironov:2009ib, Mironov:2010zs, Itoyama:2010ki, Mironov:2010ym, Morozov:2010cq, Morozov:2010uu, 
  Itoyama:2010na}.
  
  While the understanding of the AGT conjecture for a sphere 
  through the matrix model has been developed,
  a counterpart for a Riemann surface with higher genus is not yet fully understood.
  In this paper, we study the following integral:
    \bea
    Z 
     =     \int \prod_{i=1}^N dz_i \prod_{i<j} \theta_* (z_i-z_j)^{-2b^2}
           \exp \left( - \frac{b}{g_s} \sum_{j=1}^N \left( \sum_{k=1}^n 2 m_k 
           \log \theta_* (z_j-w_k) + 4 \pi i a z_j \right) \right),
           \label{torusZDF}
    \eea
  where the function $\theta_*(z)$ is defined in terms of 
  the Jacobi theta function $\theta_1(z|\tau)$
  and the Dedekind eta function $\eta(\tau)$ as 
  $\theta_* (z) \equiv q^{-1/12}\theta_1(z|\tau)/\eta(\tau)$
  with the fixed modulus $q=e^{2 \pi i \tau}$ of the torus. 
  This integral is an extension of the beta-deformed matrix model 
  corresponding to the conformal block on the sphere
  to that on the torus with $n$ punctures, 
  where $z-w$ is replaced by $\theta_*(z-w)$ \cite{DV}.
    \footnote{In \cite{Mironov:2010zs}, a slightly different integral was studied.
    Instead of introducing a factor $ 4 \pi i a z_j$ as in (\ref{torusZDF}),
    $N_1$ integrals over A-cycle and $N_2$ integrals over B-cycle are introduced in
    order to reproduce the expected number of free parameters. 
    However, it was also discussed in \cite{Mironov:2010zs}
    that their results do not completely reproduce the conformal block 
    although it is very close.}
  A closely related free field representation for toric conformal block
  is discussed in \cite{Felder:1988zp}.
 
  Although this integral cannot be written in terms of a usual matrix integral,
  it can be seen as a ``generalized matrix model'', whose ``eigenvalues'' $z_i$ live on the torus.
  Rigorously speaking, the integral is defined on the cover of the torus
  because the integrand itself is not completely doubly periodic.
  However, the positions of cuts, at which the eigenvalues are placed, are indeed doubly periodic.
  When we regard the integral variables $z_i$ as the eigenvalues,
  the product of theta functions $\prod_{i<j} \theta_* (z_i-z_j)^{-2b^2}$ can be
  regarded as a counterpart of the Vandermonde determinant.
  This form is expected from the propagator of the two-dimensional free theory on a torus.
  The parameters $m_k$ are the momenta of the vertex operators
  while $w_k$ are their insertion points.
  The parameter $a$ and the filling fractions are identified as 
  the internal momenta in the conformal block.
  The parameters in the integral are related as
    \bea
    \sum_{k=1}^n m_k + b g_s N
     =     0
    \eea
  due to the momentum conservation law.

  The AGT conjecture indicates that the $n$-point conformal block on a torus  
  is related to the $\CN=2$ superconformal $SU(2)^n$ gauge theory with a circular quiver diagram.
  The parameters $m_k$ are identified with the masses of the bifundamental hypermultiplets
  and the internal momenta of the conformal block correspond to the Coulomb moduli parameters.
  We see that such relation can be partially understood through the integral (\ref{torusZDF}) 
  in the large $N$ limit:
  On one hand, we show that the integral is derived from the perturbative calculation
  of the correlation function of the Liouville theory.
  On the other hand, we confirm that the spectral curve of the integral
  can be identified with the Seiberg-Witten curve of the quiver gauge theory.
  
  The organization of this paper is as follows.
  In section \ref{sec:Mtheory}, we review the M-theory construction 
  of the four-dimensional superconformal $SU(2)^n$ gauge theory
  and see the M-theory curve which describes the low energy effective theory of the gauge theory.
  In section \ref{sec:Liouville}, we see that the integral representation (\ref{torusZDF}) is derived 
  from the Liouville $n$-point correlation function on a torus by perturbative calculation.
  We then discuss that the proposed integral (\ref{torusZDF}) corresponds to the toric conformal block.
  In section 4, we consider the large $N$ spectral curve of the integral and identify it 
  with the curve obtained in section \ref{sec:Mtheory}.
  We conclude with discussions in section \ref{sec:conclusion}.

\section{M-theory curve of $SU(2)$ quiver gauge theory}
\label{sec:Mtheory}
\begin{figure}[t]
    \begin{center}
    \includegraphics[scale=0.7]{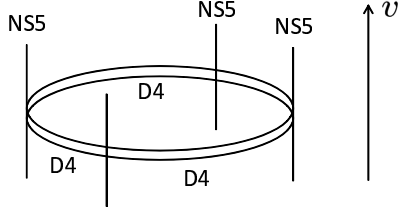}
    \caption{
    Brane configuration of the $SU(2)^4$ gauge theory with circular quiver.
    The D4-branes are suspended between the NS5-branes.}
    \label{fig:quiverT14}
    \end{center}
    \end{figure}

  In this paper, we consider $\CN=2$ superconformal $SU(2)^n$ gauge theory with circular quiver.
  This type of quiver gauge theory can be constructed as a worldvolume theory of D4-branes 
  suspended between NS5-branes in type IIA string theory \cite{Witten}, 
  as depicted in figure \ref{fig:quiverT14}.
  The D4-branes occupy the $x^{0,1,2,3}$ and $x^6$ directions 
  and the NS5-branes occupy $x^{0,1,2,3}$ and $x^{4,5}$.
  The $x^{4,5}$ coordinates are combined into the complex one $v = x^4 + i x^5$.
  The difference of the $v$-coordinates of D4-branes in 
  neighboring intervals of the NS5-branes 
  is identified with the mass parameter of the bifundamental hypermultiplet.
  The $x^6$ direction is compactified and is periodic.
  
  The curve which describes the low energy effective theory of this quiver gauge theory can be obtained 
  by the M-theory up-lift by adding periodic $x^{10}$ direction 
  and considering the hypersurface in $v, z = x^6 + ix^{10}$ space.
  Since both the $x^6$ and $x^{10}$ directions are compactified, these represent a torus $E$.
  We denote this torus by Weierstrass form
    \bea
    y^2 
     =     4 x^3 - g_2 x - g_3,
    \eea
  where these coordinates are related to $z$ in terms of  
  Weierstrass elliptic function as $x=\wp(z)$ and $y=\wp'(z)$.
  The curve in $X = \mathbb{C} \times E$ which the M5-brane wraps on 
  is identified with the Seiberg-Witten curve.
  For the $SU(2)^n$ case, the form of it is $F(x,y,v) = 0$ with
    \bea
    F(x,y,v)
     =     v^2 - f_1 (x,y) v + f_2(x,y).
    \eea
  The positions of $n$ NS5-branes are translated in M-theory to the points in the torus $E$
  in which $f_1$ and $f_2$ have simple poles \cite{Witten}.
  The residues of $f_1$ are interpreted as the mass parameters $m_k$ of the bifundamentals.
  (When we discuss poles or residues, we use the local coordinate $z=x^6+ix^{10}$.)  

  Note that the above consideration corresponds to the case 
  where the total sum of the hypermultiplet masses vanishes:
    \bea
    \sum_{k=1}^n m_k
     =      0.
    \eea
  In order to include the most generic case with non-zero sum $m$ of the masses, 
  we have to consider non-trivial bundle $X_m \rightarrow E$ such that $x^6 \rightarrow x^6 + 2 \pi R$,
  $v \rightarrow v + m$, rather than the trivial bundle $X$.
  By this choice, the constraints on $f_1$ and $f_2$ are as follows:
  let $p_k \in E$ ($k= 1, \ldots, n$) be the positions of the singularities coming from the NS5-branes.
  Then, $f_1$ and $f_2$ have order $1$ and $2$ poles respectively at, say $p_1$, and
  have simple poles at $p_2, \ldots, p_n$.
  (Note that the sum of the residues of $f_1$ should vanish 
  and therefore the residue at $p_1$ of $f_1$ is $- \sum_{k=2}^n m_k$.)
  Furthermore, the double pole of $f_2$ at $z_1$ disappears 
  when we use a good coordinate $\tilde{v} = v + \frac{m}{4} \frac{y}{x}$.

  However, as pointed out in \cite{Gaiotto}, 
  the problem can be simplified by eliminating the linear term in $v$
  and writing the curve as follows
    \bea
    v^2 
     =     \phi_2,
           \label{Mtheorycurve}
    \eea
  where we have shifted as $v \rightarrow v + \frac{f_1}{2}$ and therefore 
  $\phi_2 = \frac{f_1^2}{4} - f_2$.
  It follows from the above constraints that $\phi_2$ has double poles at $p_1, p_2, \ldots, p_n$
  and the coefficients are simply the mass parameters $m_k^2$.
  The Coulomb branch parameters are included in less singular terms:
  we can introduce $n-1$ parameters $c_k$ by adding the functions which have simple poles at $p_k$
  because the sum of the $c_k$ should vanish.
  We are also free to add the constant term in $\phi_2$.
  These correspond to $n$ Coulomb moduli parameters.
  Indeed, the most singular part of $\phi_2$ can be represented
  by $\wp(z)$ 
  and the less singular part is by elliptic $\zeta$ function: 
  $\wp = - \frac{d}{dz} \zeta$.
  Therefore, the curve is $v^2 = \phi_2$ with
    \bea
    \phi_2
     =    \sum_{k=1}^n m_k^2 \wp(z - z_k) + \sum_{k=1}^n c_k \zeta(z - z_k) + A,
    \label{SW_curve}
    \eea
  where $z_k$ are the values of the local coordinate $z = x^6 + ix^{10}$ at the points $p_k$.
  This curve can be seen as a double cover of the torus with $n$ punctures
  whose positions are $z = z_k$.
  The Seiberg-Witten one form is represented by $v dz$.

\section{From Liouville theory to generalized matrix model}
\label{sec:Liouville}
  In \cite{AGT}, it has been proposed that the $n$-point conformal block on a torus
  can be identified with the Nekrasov partition function of $\CN=2$ quiver gauge theory 
  discussed in the previous section.
  This relation for a torus was further studied in \cite{Alba:2009fp, FL}.
  The identification of the parameters of the two is as follows: 
  the momenta of the vertex operators, the internal momenta and the complex structure moduli 
  correspond, respectively, to the mass parameters of the hypermultiplets, the Coulomb moduli 
  and the gauge coupling constants in the gauge theory.
  The Liouville correlation function can be obtained
  by integrating the contribution from the three-point function
  and the conformal block over the internal momenta.
  In this section, we explain how the integral representation (\ref{torusZDF}) appears 
  from the full correlation function, based on the discussion of \cite{Goulian:1990qr, DV}.

  The $n$-point function of the Liouville theory on a torus
  is formally given by the following path integral 
    \bea
    A 
     \equiv \left< \prod_{k=1}^n e^{2 m_k\phi(w_k, \bar{w}_k)} \right>_{{\rm Liouville \,\, on}\,\, T^2}
     =     \int {\mathcal{D}} \phi (z,\bar{z}) e^{-S[\phi]} \prod_{k=1}^n e^{2m_k \phi (w_k,\bar{w}_k)},
    \eea
  where the Liouville action is given by 
    \bea
    S[\phi] 
     =     \int d^2 z \sqrt{-g} 
           \left( \frac{1}{4\pi} \partial_a \phi \partial^a \phi + \mu e^{2b \phi} \right),
    \eea
  under the flat background metric.
  We choose the insertion points $w_k$ such that they satisfy
    \bea
    \sum_{k=1}^n m_k w_k 
    =     0, 
    \label{mw0}
    \eea
  which does not break generality due to the translational invariance of the torus.

  We divide the Liouville field into the zero mode and the fluctuation
  $\phi (z,\bar{z}) = \phi_0 + \tilde{\phi} (z,\bar{z})$.
  By integrating over $\phi_0$, we obtain
    \bea
    A 
     &=&    \frac{\mu^{-\sum_{k=1}^n \frac{m_k}{b}}}{2b} \Gamma \left( \sum_{k=1}^n \frac{m_k}{b} \right)
            \int {\mathcal{D}} \tilde{\phi} (z,\bar{z}) e^{-S_0[\tilde{\phi}]} 
            \left( \int d^2 z \,\, e^{2b \tilde{\phi} (z,\bar{z})} \right) ^{-\sum_k \frac{m_k}{b}}
            \prod_{k=1}^n e^{2m_k \tilde{\phi} (w_k,\bar{w_k})} ,
    \eea
  where $S_0$ is the free scalar field action.
  When 
    \bea
    N 
     \equiv - \sum_{k=1}^{n} \frac{m_k}{b} \in \mathbb{Z}_{\geq 0}, 
    \label{mom_cons}
    \eea 
  the correlator diverges due to the factor $\Gamma(-N)$.
  Up to this divergent factor, the $n$-point correlation function $A$ is given 
  by a perturbation from the free theory 
    \bea
    A 
    &=&    \frac{\mu^{N}}{2b} \Gamma \left( -N \right) 
           \int_{T^2} \prod_{i=1}^N d^2 z_i \left\langle \prod_{i=1}^N :e^{2 b \phi (z_i, \bar{z}_i)}: 
           \,\, \prod_{k=1}^n : e^{2 m_k \phi (w_k, \bar{w}_k)} : 
           \right\rangle _{{\rm free \,\, on} \,\, T^2}.
           \label{A1}
    \eea
  The condition (\ref{mom_cons}) ensures the momentum conservation in the free theory.

  The $\ell$-point function of the free theory on a torus is given by%
  \footnote{See the review \cite{dhp} and the references therein.}
    \bea
    &&\left\langle \prod_{i=1}^{\ell} :e^{i k_i \phi (z_i, \bar{z}_i)}: \right\rangle _{{\rm free \,\, on} \,\, T^2}
     =     i C_{T^2}^X (\tau) (2\pi) \delta ( \sum _i k_i )
           \prod_{i<j} \left| \eta(\tau)^{-3} 
           \theta_1 \left( z_{ij} | \tau \right)
           \exp \left[ - \frac{\pi({\rm Im} z_{ij})^2}{\tau_2} \right] \right| ^{k_ik_j},
           \nonumber \\
    & &    \label{Npt_free}
    \eea
  where $z_{ij} \equiv z_i-z_j$, $\tau$ is the modulus of the torus, $\tau_2$ is its imaginary part, 
  and $C_{T^2}^X = (4 \pi^2 \tau_2)^{-\frac{1}{2}} |\eta (\tau)|^{-2}$.
  It factorizes into holomorphic and anti-holomorphic parts 
  by introducing an additional integral as \cite{VV, DVV, dhp}
    \bea
    &&\left\langle \prod_{i=1}^{\ell} :e^{i k_i \phi (z_i, \bar{z}_i)}: \right\rangle 
    _{{\rm free \,\, on} \,\, T^2}
    \cr
    &&=     2 i |\eta (\tau)|^{-2} \delta ( \sum _i k_i )  
            \int^{\infty}_{-\infty} da \left| \left( \prod_{i<j} 
           \left( \frac{\theta_1 \left( z_{ij} | \tau \right)}{\eta(\tau)^{3}} \right) 
           ^{\frac{k_ik_j}{2}} \right) q^{a^2}
           \exp \left( - 2 \pi i \sum_{j=1}^{\ell} k_j z_j a \right)
           \right| ^2 ,
           \label{factorized}
    \eea
  where $q=\exp(2 \pi i \tau)$.
  We recover (\ref{Npt_free}) by explicitly carrying out the Gaussian integral over $a$.
  This expression is more suitable for associating the 
  Liouville $n$-point function to the holomorphic generalized matrix model.

  Here, we introduce 
  \bea
  \theta_* (z) 
   \equiv q^{-1/12} \frac{\theta_1(z|\tau)}{\eta(\tau)}
   =     2 \sin(\pi z) \prod_{m=1}^{\infty} (1-e^{2 \pi i z}q^m) (1-e^{-2 \pi i z}q^m)
         \label{def_theta}
  \eea
  where $q = \exp(2 \pi i \tau)$ is fixed, for later convenience.
  Using the explicit expression (\ref{factorized}) for (\ref{A1}), 
  we find that the $n$-point function $A$ of the Liouville theory reduces to the following integral 
    \bea
    A 
    &=&    C (\tau,m_k,b) \prod_{1\le k<l \le n} 
           \left| \theta_* ( w_{kl} )\right| ^{-4m_k m_l}
           \int^{i\infty}_{-i\infty} da \,\, |q|^{-2a^2} \int_{T^2} \prod_{i=1}^N d^2 z_i 
    \cr
    & &    \qquad \left| \exp \left[ 
           - 2b \sum_{i=1}^N \sum_{k=1}^n m_k \log \theta_* \left( z_{i}-w_{k} \right) ^{} 
           - 2b^2 \sum_{1 \le i<j \le N} \log \theta_* ( z_{ij} ) 
           - 4 \pi i b a  \sum_{i=1}^N z_i 
           \right] \right| ^2,
           \label{Amp_ctdt}
    \eea
  where $w_{ij} \equiv w_i-w_j$, and we have introduced the factor in front of the $z$ integral as
    \bea
    C(\tau,m_k,b) 
    \equiv \frac{\mu^{N}\Gamma(-N)}{b} \delta(0) |\eta (\tau)|^{-2} 
           |q^{-1/24} \eta(\tau)| ^{ - 4 \sum_k m_k{}^2 - 4b^2 N }.
             \label{C}
    \eea 

  The discussion above is valid even for finite $N$.
  However, it is not straightforward to factorize the integrals over the torus into
  holomorphic and anti-holomorphic integrals for generic $N$.
  In order to proceed to the next step, 
  we evaluate the integral (\ref{Amp_ctdt}) in the large $N$ limit.
  From the momentum conservation (\ref{mom_cons}),
  we see that $m_k = {\cal O} (N)$ as $b = {\cal O} (1)$.
  Although the integral variable $a$ runs over the whole range of imaginary numbers,
  dependence on $a$ of the exponent appears when $a={\cal O} (N)$.
  We see that all the three terms in the exponent in (\ref{Amp_ctdt}) are all ${\cal O}(N^2)$.
  In the large $N$ limit, the integral (\ref{Amp_ctdt})
  is evaluated at the critical points of the exponent of the integrand.
  The conditions for the criticality of the exponent are given by
    \bea
    \sum_{k=1}^n m_k \frac{\theta_*'(z_i-w_k)}{\theta_*(z_i-w_k)}
    + b \sum_{j \neq i}\frac{\theta_*'(z_{ij})}{\theta_*(z_{ij})}
    + 2 \pi i a = 0 
    \label{ext_cond}
    \eea
  where $\theta_*'(z) \equiv \partial_z \theta_*(z)$.
  The conditions obtained from the $\bar{z}_i$-derivatives
  are just the complex conjugate of (\ref{ext_cond}).
  It is remarkable that the conditions for criticality are separated into
  holomorphic and anti-holomorphic equations,
  which indicates that the integrals over the torus in (\ref{Amp_ctdt}) can be factorized
  into holomorphic and anti-holomorphic integrals in the large $N$ limit.

  If we do not have the second term in (\ref{ext_cond}), 
  we have $n$ possible critical points for each variable $z_i$,
  assuming that the parameters $m_k$ are generic.
  We expect that the $n$ critical points are ``diffused'' to form line segments due to the second term,
  similarly to the case of the usual large $N$ matrix model.
  Then, the solutions of (\ref{ext_cond}) are labelled by the filling fraction $\nu_k = b g_s N_k$,
  in which $N_k$ out of $N$ variables $z_i$ take the value on the $k$-th line segment.
  Here, we have introduced the parameter $g_s$, where $g_s N$ is finite in the large $N$ limit, 
  for later convenience.

  We define the contributions from the solution labelled by $\{ \nu_k \}$ to the holomorphic integral as 
    \bea
    Z(q,w_k,m_k,a,\nu_k)
    &\equiv& 
           \int \prod_{i=1}^N dz_i 
           \prod_{1 \le i < j \le N} \theta_* (z_{ij})^{-2b^2}
           \nonumber \\
    & &    ~~~~~~
           \exp \left( - \frac{b}{g_s} \sum_{i=1}^N 
           \left( \sum_{k=1}^n 2 m_k \log \theta_* (z_i-w_k) + 4 \pi i a z_i \right) \right),
           \label{holo}
    \eea
  where we have rescaled the parameters as $m_k \rightarrow m_k/g_s$ and $a \rightarrow a/g_s$. 
  The paths of the integrals are defined such that 
  only the solution of (\ref{ext_cond}) labelled by the fixed filling fractions $\{ \nu_k \}$
  contributes to the integrals.
  By regarding the factor $\prod_{1 \le i < j \le N} \theta_* (z_{ij})^{-2b^2}$
  as the generalization of the Vandermonde determinant,
  we see that the holomorphic integral (\ref{holo}) is ``the generalized matrix model'' with the action
    \bea
    W(z) =  \sum_{k=1}^n 2 m_k \log \theta_* (z-w_k) + 4 \pi i a z.
    \label{c_action}
    \eea

  The integral in (\ref{Amp_ctdt}) is then obtained by integrating (\ref{holo}) 
  and its complex conjugate over the filling fractions.
  Thus, in the large $N$ limit, the $n$-point function $A$ of the Liouville theory can be written as
    \bea
    A 
    &=&    \int^{i\infty}_{-i\infty} da \int d \nu_1 \cdots d \nu_{n-1} 
           \cr
           ~~~
    & &    \left| 
           \left( q^{-1/24} \eta(\tau) \right) ^{ - 2 \sum_k m_k{}^2 / g_s^2}
           \left( \prod_{1\le k<l \le n} \theta_* (w_{kl}) ^{- 2 m_k m_l / g_s^2} \right)
           q^{- a^2/g_s^2} Z(q,w_k,m_k,a,\nu_k) 
           \right|^2 \!\!\!\! ,
           \label{FinalA}
    \eea
  where we have used that $C (\tau,m_k,b)$ is approximated as
  $\left| q^{-1/24} \eta(\tau)\right| ^{-4\sum_k m_k{}^2/g_s^2}$.

  The total $n$ parameters, 
  $a$ and the independent filling fractions $\nu_1, \cdots, \nu_{n-1}$,
  can be identified with the $n$ internal momenta $\alpha_1, \cdots , \alpha_n$
  in the conformal block $\CB(q,w_k,m_k,\alpha_k)$ \cite{DV}.
  Under this identification, we see from (\ref{FinalA}) that  
  $Z(q,w_k,m_k,a,\nu_k)$ corresponds to the conformal block 
  as well as the holomorphic contribution from the three-point functions.

  In the next section, we relate the generalized matrix model (\ref{holo})
  with the $\CN=2$ quiver gauge theory in the previous section.
  Before going into it, let us see the remaining parts in (\ref{FinalA}) here.
  As discussed in \cite{AGT}, the gauge coupling constants
  $q_i = \exp (2 \pi i \tau_i)$ ($i=1,\cdots, n$) of the $SU(2)^n$ quiver gauge theory is related to 
  the modulus $q=\exp(2 \pi i \tau)$ of the torus and 
  the insertion points $w_i$ of the $n$-point function of the Liouville theory as
    \bea
    q_1 
     =     e^{2 \pi i(w_1-w_2)}, \quad
    q_2 
     =     e^{2 \pi i(w_2-w_3)}, \quad \cdots, \quad 
    q_{n-1}
     =     e^{2 \pi i(w_{n-1}-w_{n})}, \quad
    q_1 q_2 \cdots q_n
     =     q.
    \eea
  Using (\ref{def_theta}) and the definition for the Dedekind eta function
  $
  q^{-1/24} \eta(\tau) = \prod_{m=1}^{\infty} (1-q^m) ,
  $
  the factor in front of $Z(q,w_k,m_k,a,\nu_k)$ in (\ref{FinalA})
  is rewritten in terms of $q_i$ as 
    \bea
    && \left( q^{-1/24} \eta(\tau) \right) ^{ - 2 \sum_k m_k{}^2 / g_s^2}
    \prod_{1\le k<l \le n} \theta_* (w_{kl}) ^{- 2 m_k m_l / g_s^2}
    \cr
    && =    \prod_{k<l} \left( (i q_k \cdots q_{l-1})^{m_k m_l/g_s^2} \right)
            \prod_{i=1}^n \prod_{k=0}^{\infty} 
            \left( 1 - q_i q_{i+1} \cdots q_{i+k}\right) ^{-2m_{i}m_{i+k+1}/g_s^2}
    \eea
  where the subscripts of $m_i$ and $q_i$ are considered modulo $n$.
  This factor has a close form as the $U(1)$ factor, 
  which is explicitly written in \cite{AGT}.
  Furthermore, the factor $q^{- a^2/g_s^2}$ in (\ref{FinalA}) corresponds to 
  the tree level contribution to the prepotential.
  
  In this section, we have derived
  the generalized matrix model from the Liouville correlation function
  by explicitly carrying out the perturbative calculation.
  In \cite{DV}, it was discussed that the action (\ref{c_action}) 
  of the generalized matrix model is also
  expected from a geometrical argument in topological string theory 
  in the context of the AGT conjecture.
 
\section{Spectral curve of the generalized matrix model}

  In this section, we derive the spectral curve of the generalized matrix model,
  which is introduced in the previous section:
    \bea
    \exp \left( - \frac{1}{g_s^2} {\cal F} \right)
     =    \int \prod_{i=1}^N d\lambda_i 
          \exp \left( 
          - \frac{b}{g_s} \sum_{i=1}^N W(\lambda_i) 
          + \sum_{i < j} \log \left( \theta_* (\lambda_i - \lambda_j) \right)^{-2b^2}
          \right) ,
    \label{GMM}
    \eea
  where $W(\lambda)$ is given by (\ref{c_action}).
  As stated previously, the paths of the integrals are determined
  such that they realize the given filling fractions.
  For generic parameters, the action $W(\lambda)$ has $n$ critical points.
  In order to make the discussion as generic as possible,
  we will calculate it with a generic action with $n$ critical points 
  and, at the final stage, substitute its specific form (\ref{c_action}).
  We will see that the Seiberg-Witten curve (\ref{Mtheorycurve}) with (\ref{SW_curve}) 
  appears as the spectral curve of this generalized large $N$ matrix model.

  In the large $N$ limit, the problem of the integration in (\ref{GMM}) 
  reduces to calculation of the critical point of its exponent. 
  The prepotential is given by
    \bea
    - \frac{1}{g_s^2} {\cal F}
     =     - \frac{b}{g_s} \sum_i W(\lambda_i) 
           - \sum_{i \neq j} b^2 \log \theta_* (\lambda_i - \lambda_j),
    \eea
  where each eigenvalue satisfies the condition of criticality
    \bea
    0 =     \frac{1}{g_s} W'(\lambda_i) 
            + 2b \sum_{j (\neq i)} 
              \frac{\theta_*' (\lambda_i - \lambda_j)}{\theta_* (\lambda_i - \lambda_j)} .
    \label{ex_con}
    \eea

  It is natural to assume that the eigenvalues 
  are distributed around the critical points of $W(\lambda)$ but in the form of line segment,
  similarly to the usual matrix model.
  We denote these line segments as $C_k$ where $k=1,\cdots n$.
  We do not assume that $C_k$ are on a real axis.
  However, we assume that $C_k$ do not include the singular points $w_\ell$,
  at which the action $W(\lambda)$ diverges.
  Suppose that $N_k$ eigenvalues are on the line segment $C_k$,
  where $N_k$ satisfies $\sum_{k=1}^n N_k = N$.

  Here, we introduce the density of eigenvalues $\rho(\lambda)$ 
  which has non-zero value only on the line segment $C_k$.
  Outside of these regions, we define that $\rho (\lambda) = 0$.
  The density of eigenvalues is normalized as
  $\int_{C_k} d \lambda \rho (\lambda) = b g_s N_k \equiv \nu_k$.
  Using the variables introduced above, the prepotential and the condition
  for criticality are written as
    \bea
    &&{\cal F} 
     =     \int_{\sum_k C_k} d\lambda \rho (\lambda) W(\lambda)
           + \int_{\sum_k C_k} d\lambda' \int_{\sum_k C_k} d\lambda \rho (\lambda) \rho (\lambda') 
           \log \theta_*(\lambda-\lambda')
           \label{prepot}, \\
    && 0 
     =     W(\lambda)' 
           + 2 \int_{\sum_k C_k} d \lambda' \rho (\lambda') 
           \frac{\theta_*' (\lambda - \lambda')}{\theta_* (\lambda - \lambda')} ,
           \label{cond_extrem}
    \eea
  respectively.

  In order to solve this, we define the resolvent as 
    \bea
    R(z) 
    \equiv \int_{\sum_k C_k} d\lambda \, \rho (\lambda) 
           \frac{\theta_*' (z - \lambda)}{\theta_* (z - \lambda)}.
           \label{resolvent}
    \eea
  Since the behavior of the theta function around the zeros is 
  $\theta_*(\varepsilon) \sim C \varepsilon + {\cal O} (\varepsilon^3)$, 
  $\theta_*'(\varepsilon) \sim C + {\cal O} (\varepsilon^2)$ 
  for small $\varepsilon$,
  the structure of the singularity is similar to that of the usual matrix model
  if we focus on the fundamental region of the torus.
  The resolvent has cuts at the line segments $C_k$.
  Also, the filling fractions are obtained by integrating the resolvent along the cuts as
    \bea
    \nu_k 
     =     \frac{1}{2\pi i} \oint_{C_k} dz R(z).
           \label{filling}
    \eea

  The significant difference from the usual hermitian matrix model is 
  that the resolvent has pseudo periodicity due to the theta function.
  Using the identity
    \bea
    \frac{d}{dz} \log \theta_*(z+m+n\tau,\tau) 
     =     - 2 \pi i n + \frac{d}{dz} \log \theta_* (z,\tau),
    \label{theta_cyclic}
    \eea
  which holds for arbitrary integer $m$ and $n$,
  the pseudo periodicity of the resolvent is shown to be given as
    \bea
    R(z+m+n\tau) 
     =     R(z) - 2 \pi i n b g_s N.
           \label{cyclic}
    \eea
  We see that the resolvent is completely periodic for the $A$-cycle, 
  which is parallel to the real axis,
  but some constant is added for the $B$-cycle. 
  This pseudo periodicity gives great restriction on the possible form of the resolvent below.

  On the line segments $C_k$, the resolvent is expected to behave as
    \bea
    &&R(z + i \varepsilon e^{i\varphi(z)}) + R(z - i \varepsilon e^{i\varphi(z)}) 
     =     2 \int_{\sum_k C_k} d \lambda' \rho(\lambda') 
           \frac{\theta_*' (z - \lambda')}{\theta_* (z - \lambda')}
     =     -  W ' (z),
           \label{prin} \\
    &&R(z + i \varepsilon e^{i\varphi(z)}) - R(z - i \varepsilon e^{i\varphi(z)}) 
     =     \oint_{z} d \lambda' \rho(\lambda') 
           \frac{\theta_*' (z - \lambda')}{\theta_* (z - \lambda')}
     =     - 2 \pi i \rho (z),
           \label{cut} 
    \eea
  where we take real number $\varepsilon$ infinitely small 
  and $\varphi(z)$ is properly defined such that 
  $z + i \varepsilon e^{i\varphi(z)}$ or $z - i \varepsilon e^{i\varphi(z)}$
  does not go across the cuts $C_k$ when $z$ moves along $C_k$. 
  The integral in (\ref{prin}) is principal integration,
  which is given as an average of integral 
  along the path above the singularity and that below the singularity.
  The resolvent should be determined such that (\ref{prin}) and (\ref{cut}) 
  are satisfied for $z \in C_k$,
  together with the periodicity (\ref{cyclic}).

  A candidate of the solution for (\ref{prin}) is
    \bea
    R_0(z) 
     =     - \frac{1}{2} W'(z).
    \eea
  However, it does not reproduce the correct structure of singularity expressed in (\ref{cut}).
  We need singular contributions:
    \bea
    R(z) 
     =     R_0(z) + R(z)_{\rm sing}, 
           \label{R_def}
    \eea
  where (\ref{prin}) and (\ref{cut}) impose
    \bea
    R(z + i \varepsilon e^{i\varphi(z)})_{\rm sing} + R(z - i \varepsilon e^{i\varphi(z)})_{\rm sing}
    &=&     0.
            \label{sing_prin}
    \\
    R(z + i \varepsilon e^{i\varphi(z)})_{\rm sing} - R(z - i \varepsilon e^{i\varphi(z)})_{\rm sing}
    &=&     - 2 \pi i \rho (z).
            \label{sing_cut} 
    \eea

  The above discussion is valid for generic action $W(z)$.
  In the following, we use the concrete form (\ref{c_action}) to determine the resolvent $ R(z) $.
  Since $R_0(z)$ is given by
    \bea
    R_0(z)
     =     - \sum_{k=1}^n m_k \frac{d}{d z} \log \theta_* (z - w_k) - 2 \pi i a,
           \label{R0}
    \eea
  we see that its pseudo periodicity is given by
    \bea
    R_0(z + m + n \tau) 
     =     R_0(z) - 2 \pi i n b g_s N,
           \label{W_cyclic}
    \eea
  where we used the identity (\ref{theta_cyclic}) and the momentum conservation (\ref{mom_cons}).
  (Note that the parameters $m_k$ and $a$ have been rescaled.) 
  Remarkably, the pseudo periodicity (\ref{W_cyclic}) for $R_0(z)$ 
  exactly agrees with that of the resolvent $R(z)$ in (\ref{cyclic}).
  Therefore, we find that $R(z)_{\rm sing}$ is exactly doubly periodic.

  On one hand, we find that the resolvent $R(z)$ has no singularity 
  except for the cuts in the regions $C_k$ from the definition (\ref{resolvent}).
  On the other hand, we see from the explicit form (\ref{R0})
  that $R_0(z)$ has simple poles at $z=w_k+m+n\tau$ and their residues are $-m_k$. 
  Therefore, $R(z)_{\rm sing}$ must have both the cuts in the regions $C_k$
  and simple poles with residues $m_k$ at $z=w_k+m+n\tau$. 

  From (\ref{sing_prin}), we see that the sign of $R(z)_{\rm sing}$ changes 
  when crossing the cut.
  Thus, $R(z)_{\rm sing}^2$ has no such discontinuity and all the cuts disappear.
  Since $R(z)_{\rm sing}^2$ has double poles at $z=w_k+m+n\tau$,
  the Laurent expansion of $R(z)_{\rm sing}^2$ at $z=w_k$ has
  not only $(z-w_k)^{-2}$ terms but also $(z-w_k)^{-1}$ terms due to contact terms.
  We denote the coefficients of such terms as $c_k$.
  Such doubly periodic function is uniquely determined, up to adding a constant,
  as a linear combination of the Weierstrass $\zeta$ function,
  which has one simple pole in the fundamental region,
  and of the Weierstrass $\wp$ function,
  which has one double pole in the fundamental region. 
  This uniqueness is followed from the Liouville theorem,
  which ensures that a function holomorphic on the whole complex plane is only the constant function.
  In other words, we cannot add non-trivial function to the doubly periodic function
  without changing its structure of singularity.

  From the discussion above, we have derived that $R(z)_{\rm sing}{}^2$ must be of the form
    \bea
    \left( R(z) + \frac{1}{2} W'(z) \right)^2
     =     R(z)_{\rm sing}{}^2 
     =     \sum_{k=1}^n m_k^2 \wp (z-w_k) + \sum_{k=1}^n c_k \zeta (z-w_k) - C.
           \label{curve}
    \eea
  where the coefficients $c_k$ satisfy the condition
    \bea
    \sum_{k=1}^n c_k
      =     0
            \label{cond_ck}
    \eea
  so that (\ref{curve}) is doubly periodic.
  This is the spectral curve of the generalized matrix model.
  This spectral curve (\ref{curve}) coincides with the form of 
  the Seiberg-Witten curve (\ref{Mtheorycurve}) with (\ref{SW_curve}).

\subsection{One-point function}
  In the following, we concentrate on the case of one-point function
  and determine the constant in the spectral curve under the approximation that $|a|$ is large enough.
  The spectral curve (\ref{curve}) for $n=1$ reduces to
    \bea
    R(z)_{\rm sing}{}^2 
     =     m^2 \wp (z) - C,
           \label{1pt_curve}
    \eea
  where we used the condition (\ref{cond_ck}).
  From (\ref{sing_cut}), the density of eigenvalue is written as
    \bea
    \rho(\lambda) 
     =     (\pm) \frac{1}{\pi i} \sqrt{ m^2 \wp(\lambda) - C},
           \label{1pt_rho}
    \eea
  where the sign should be determined so that
  $
  - i \rho(\lambda) d\lambda  
  $
  is positive real on the integral path $C_1$.

  The equation of motion (\ref{cond_extrem}) for $n=1$ reduces to
    \bea
    m \frac{\theta_*' (\lambda)}{\theta_* (\lambda)} + 2 \pi i a
    + \int_{C_1} d\lambda' \rho(\lambda') 
    \frac{\theta_*' (\lambda - \lambda')}{\theta_* (\lambda - \lambda')}
     =     0.
           \label{1pt_eom}
    \eea
  Now, suppose that $ia \gg 1$.
  In order that the eigenvalues satisfy the equation of motion, 
  at least either the first or the third term of (\ref{1pt_eom}) must be large enough 
  to compensate the second term.
  The first term becomes large if an eigenvalue is placed close to the singularity $\lambda = 0$.
  The third term becomes large if the distribution $\rho(\lambda')$ takes large value around the pole
  $\lambda' = \lambda$ of the integrand.
  Since $\rho(\lambda')$ is constrained by the normalization condition 
    \bea
    \int _{C_1} d\lambda' \rho (\lambda') 
     =     b g_s N,
           \label{1pt_norm}
    \eea
  $\rho(\lambda')$ must take large value only around the region close to $\lambda' = \lambda$
  and rapidly decrease as it goes apart in this case.
  However, since $\rho (\lambda')$ is given by (\ref{1pt_rho}), 
  this occurs only if $\lambda$ is close to the singularity $\lambda = 0$.
  Thus, in any case, every eigenvalue $\lambda$ 
  must be distributed very close to the singularity $\lambda = 0$.
  In other words, the cut $C_1$ of the resolvent is placed very close to the singularity $z=0$
  and its length is very short.

\begin{figure}
    \begin{minipage}{0.5\hsize}
    \centering
    \includegraphics[width=5cm]{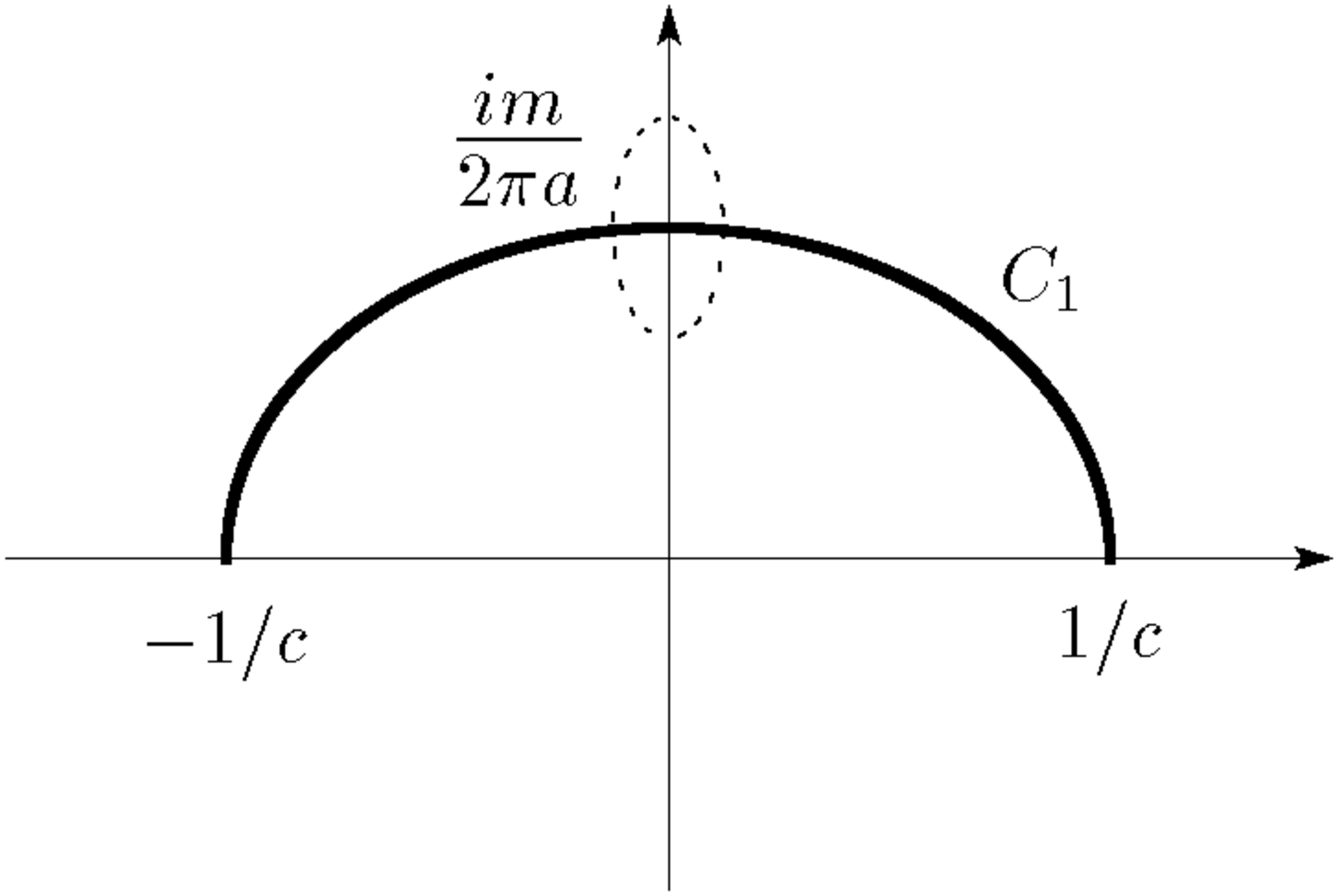}
    \caption{Path $C_1$}
    \label{zu}
    \end{minipage}
    \begin{minipage}{0.5\hsize}
    \centering
    \includegraphics[width=5cm]{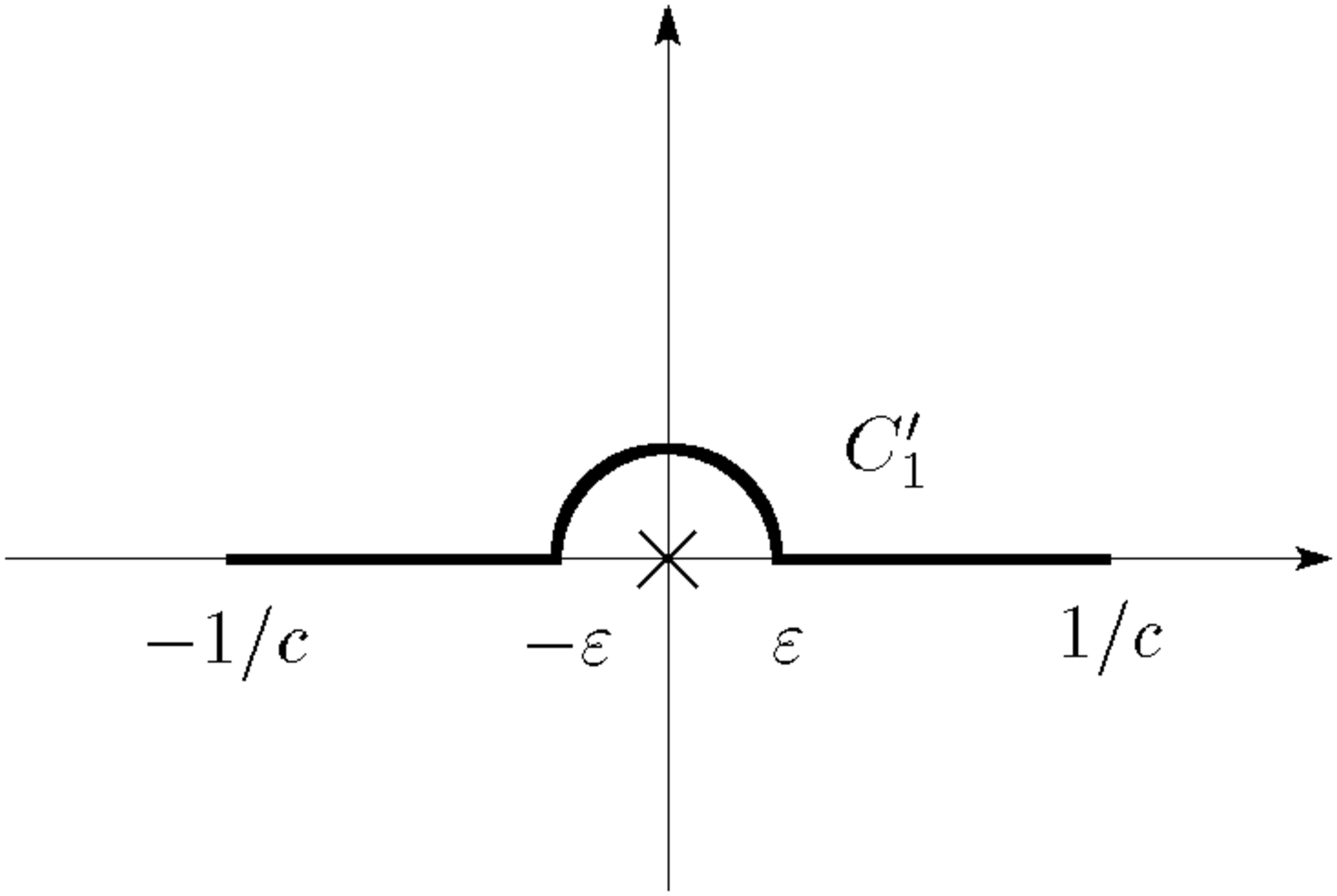}
    \caption{Path $C_1'$}
    \label{zu2}
    \end{minipage}
    \end{figure}

  When $\lambda \sim 0$, the eigenvalue density (\ref{1pt_rho}) can be approximated as
    \bea
    \rho 
     \sim  (\pm) \frac{m}{\pi i} \sqrt{ \frac{1}{\lambda^2} - c^2},
           \label{app_rho}
    \eea
  where we put $C \equiv m^2 c^2$ for future convenience.
  Therefore, in this approximation, the endpoints $a_1$, $b_1$ of the cut $C_1$ 
    ($\rho (a_1) 
     =     \rho (b_1) 
     =     0$)
  are
    \bea
    a_1,b_1 
     \sim  \pm \frac{1}{c}.
    \eea
  Also, the equation of motion (\ref{1pt_eom}) can be approximated as
    \bea
    \frac{m}{\lambda} + 2 \pi i a + \int_{C_1} d\lambda' \frac{\rho(\lambda')}{\lambda - \lambda'}
     =     0.
           \label{app_eom}
    \eea
  Since the solution of the classical equation of motion is given by 
    \bea
    \lambda 
     =     \frac{im}{2 \pi a},
    \eea
  it is expected that the cut $C_1$ is placed around this point.
  Taking account the symmetry, most natural possibility is that
  the endpoints $a_1, b_1 = \pm 1/c$ are on the real axis and the path $C_1$ goes 
  beyond the origin as in figure \ref{zu}. 

  For later calculation, we analytically continue the eigenvalue density $\rho(\lambda)$,
  which has non-zero value only on the cut,
  to the whole complex plane and we denote it as $\tilde{\rho}(z)$.
  However, if we define $\tilde{\rho}(z)$ by the function in (\ref{app_rho}) 
  under the usual convention that the function $f(z)=\sqrt{z}$ has cut on the negative real axis,
  it changes the sign across the imaginary axis.
  Since the original eigenvalue function $\rho(\lambda)$ does not have such discontinuity on $C_1$,
  we rewrite (\ref{app_rho}) and define $\tilde{\rho}(z)$ as 
    \bea
    \tilde{\rho} (z)
     \equiv \frac{m}{\pi i z} \sqrt{1 - (cz)^2},
           \label{good_cut}
    \eea
  so that it is smooth on the path $C_1$.
  The sign was determined so that $-i\tilde{\rho}(z)$ is positive at $z \sim im/a$.
  
  At this stage, let us check the normalization condition (\ref{1pt_norm}).
  By using the analytically continued function (\ref{good_cut}),
  we can change the path of the integral from $C_1$ to $C_1'$ as in figure \ref{zu2}
  because there are no cuts or singularity between the path $C_1$ and $C_1'$.
  Then, we can explicitly calculate the integral as
    \bea
    \int _{C_1} d\lambda \rho (\lambda)
     &=&  \int_{-1/c}^{\varepsilon} dz \tilde{\rho} (z)
          + \int_{z=\varepsilon e^{i\theta}, \theta:\pi \to 0} dz \tilde{\rho} (z)
          + \int_{\varepsilon}^{1/c} dz \tilde{\rho} (z)
    \cr
     &=& \int_{0}^{1/c} dz [ \tilde{\rho} (z) + \tilde{\rho} (-z)]
         - \frac{1}{2} \times 2 \pi i \, {\rm Res}_{z=0} \,\, \tilde{\rho} (z)
    \cr
     &=& - m .
    \eea
  Thus, taking account the momentum conservation 
  $
  b g_s N + m = 0,
  $
  we find that the normalization (\ref{1pt_norm}) is satisfied for arbitrary $c$ 
  and imposes no constraints.

  We now use the equation of motion (\ref{app_eom}) at $\lambda = 1/c$, 
  which is the endpoint of the cut $C_1$.
    \bea
    cm + 2 \pi i a + \int_{C_1} d\lambda' \frac{\rho(\lambda')}{1/c - \lambda'}
     =     0. 
    \eea
  By changing the path of the integral similarly to the previous calculation 
  and by using (\ref{good_cut}),
  we find that the integral in the equations of motion can be explicitly calculated to give
    \bea
    \int_{C_1} d\lambda' \frac{\rho(\lambda')}{1/c - \lambda'}
     =     - (1+i) mc.
    \eea
  Thus, we have determined the coefficient $c$ as
    \bea
    c 
     =     \frac{2 \pi a}{m},
    \eea
  which indicates $C = (2 \pi a) ^2$.
  Thus, the spectral curve (\ref{1pt_curve}) is 
    \bea
    R(z)_{\rm sing}  ^2 
     =    m^2 \wp (z) - (2 \pi a)^2.
    \eea
  
  In the limit $|a| \gg |m|$, it is straightforward to check that
    \bea
    \int_{\beta} R(z)_{\rm sing} 
     =     2 \pi i m, 
    ~~~~~~
    \int_{\gamma} R(z)_{\rm sing} 
     =     2 \pi i a,
    \eea
  where $\beta$ is the small cycle around the origin while $\gamma$
  is the $A$-cycle of the torus.
  These are the expected property, which the Seiberg-Witten curve should satisfy.
  Thus, we have shown that the spectral curve of the generalized matrix model 
  coincides with the Seiberg-Witten curve including the constant term in this limit. 

\section{Conclusion and discussion}
\label{sec:conclusion}

  In this paper, we have studied the generalized matrix model,
  which is a Dotsenko-Fateev type integral representation of the toric conformal block.
  This generalized matrix model is shown to be naturally derived from the perturbative calculation
  of the $n$-point function of the Liouville theory on the torus.

  We have shown that the Seiberg-Witten curve of the corresponding gauge theory 
  is derived from the generalized matrix model as a spectral curve.
  We have confirmed that 
  the constants of the spectral curve of the generalized matrix model with $n=1$ agrees with 
  those of the Seiberg-Witten curve for $\CN=2^*$ theory in a large internal momentum limit.
  Our results suggest that the AGT relation between the Liouville theory on a torus and 
  the four dimensional ${\cal N}=2$ quiver gauge theory 
  can be understood through the generalized matrix model.

  One of the future directions is to determine the undetermined constants 
  in the spectral curve in more generically.
  Then, we go on to the next step to calculate the prepotential
  and compare the results with the conformal block or with the Nekrasov partition function.

  Other direction is to consider a finite $N$ correction.
  We have seen the large $N$ limit has played an
  essential role to derive the generalized matrix model from the Liouville $n$-point function.
  Thus, such an extension seems to be quite non-trivial.
  In a sphere case, it is known that the matrix expression for 
  the conformal block is valid even for finite $N$ \cite{Fujita:2009gf} -- \cite{Itoyama:2010na}.
  So, it would be quite interesting to study whether such an expression is possible 
  also for the torus case. 
  In particular, the derivation of the loop equation would strongly help 
  to calculate finite $N$ partition function \cite{Ambjorn:1992gw, Akemann:1996zr, Eynard:2007kz}.
  
  The generalized matrix model which we have considered in this paper can be seen 
  as an elliptic extension of the Selberg integral.
  The properties of this have been studied in \cite{Felder, Spiridonov}.
  It would be interesting to study the generalized matrix model from this direction.

\section*{Acknowledgements}
  We would like to thank Kazuo Hosomichi for reading our paper carefully
  and giving some useful comments. 
  We also would like to thank Tohru Eguchi, Sylvain Ribault and Masato Taki for useful discussions and comments.
  Research of K.M.~is supported in part by JSPS Bilateral Joint Projects (JSPS-RFBR collaboration).
  F.Y.~is supported by the William Hodge Fellowship.

\end{document}